\documentstyle[12pt]{article}
\begin{document}
\title {HOW UNSTABLE ARE FUNDAMENTAL QUANTUM
SUPERMEMBRANES?}
\author {MICHIO KAKU}
\date {Physics Dept., City College of New York\\
New York, N.Y. 10031, USA}
\maketitle
\begin {center}
\Large {ABSTRACT}
\end {center}

\section{Quantum Supermembranes}
String duality, pioneered by Kikkawa and Yamasaki$^{1,2}$,
represents an enormous advance in our understanding of string
physics. For the first time, we can peer into the non-perturbative
region of certain string theories and settle questions which have
dogged the field since its very inception$^{3-8}$.

In particular, solitonic $p$-branes are necessary to
complete our understanding of BPS-saturated states.
An eleven dimensional \lq\lq M-theory," in fact, must be able to
incorporate both strings and solitonic membranes.
Because these solitonic membranes likely have finite
thickness, they are probably stable.

By contrast, fundamental quantum supermembranes are thought
to have serious problems. Besides the fact that they are
highly non-linear (and hence their spectrum is impossible
to calculate exactly), they also have several serious
physical diseases:

\noindent (a) the world volume action is not renormalizable

\noindent (b) the theory has no dilaton, so a standard KSV
type perturbation theory is not possible

\noindent (c) the theory is thought to be
unstable; for string-like configurations,
the zero-point energy of the Hamiltonian may be zero$^9$.

The first problem means that an infinite number of counter-terms
must be added to the world volume action to render it finite.
However, perhaps these counter-terms simply represent the
infinite number of background fields corresponding to 
excitations of the supermembrane. So having an infinite
number of counter-terms is by itself not necessarily
fatal.

The second problem is also not necessarily fatal if a new
mechanism is found for interacting membranes, other than
the standard dilaton formulation. Since we do not know how
supermembranes split apart (or even if they do), it is premature
to discount them on the basis of interactions.

The third problem is more serious, since it goes to the heart of
whether fundamental quantum supermembranes are stable or not.

Previously, in ref. 9, this question was addressed by
approximating the supermembrane action$^{10,11}$ by a SU(n)
super Yang-Mills theory as $n \rightarrow \infty$. For finite
$n$, this amounts to a convenient regulator for the theory.
Although this proof is rather convincing, it depends on whether
the $n\rightarrow\infty$ limit is singular or not.
Perhaps there are regularization-dependent factors which
enter into the picture in this delicate limit.

In this paper, we will try to address the question directly,
whether the continuum theory is stable or not. By analyzing
the continuum theory, we have a much more intuitive grasp
of precisely where the problems may lie and where the potential
infinities may occur.  We will follow the basic outline of
ref. 9, but adapt their calculation for our purposes.

And second, at the end of this paper, we present some rough
speculations
about how unstable membranes may still be made
into a physical theory.

We begin with the action for the membrane, which is given by:

\begin {equation}
S = S _ 1+ S _ 2
\end {equation}
where $S _ 1 $ is the usual determinant defined over a world volume:

\begin {equation}
S_ 1 = - T \int d ^ 3 \sigma \, \sqrt { -M} ;
\,\,\,
M _ { ij} = \Pi _ i ^ \mu \Pi _ j ^ \nu \eta _ { \mu \nu }
\end {equation}
where:
\begin {equation}
\Pi _ i ^ \mu = \partial _ i X ^ \mu - i \bar \theta \Gamma ^ \mu
\partial _ i \theta
\end {equation}
and $S _ 2$ is a Wess-Zumino term$^{10,11}$:

\begin {equation}
S _ 2= - T\int d ^ 3 \sigma \,
\left [
{ 1 \over 2 }
\epsilon ^ { ijk} 
\bar \theta \Gamma _ { \mu \nu } \partial _ i \theta
\left ( \Pi _ j ^ \mu \Pi _ k ^ \nu + i
\Pi _ j ^ \mu \bar \theta \Gamma ^ \nu \partial _ k \theta
- { 1 \over 3 }
\bar \theta
\Gamma ^ \mu \partial _ j \theta \bar \theta \Gamma ^ \nu
x\partial _ k \theta \right ) \right ]
\end {equation}
where $i =  1,2,3$ represents the three world volume indices
of the membrane. Two of them, $\sigma _ 1$ and $\sigma _ 2$,
represent the co-ordinates of the surface, and $\sigma _ 3 =\tau$
represents the time-like direction. The Greek symbols represent
11 dimensional Lorentz indices. $\Gamma ^\mu$ are the usual
Dirac matrices in 11 dimensions. $X ^ \mu$ is the co-ordinate
of the membrane, and $\theta$ is a Majorana spinor with
32 real components.

This action is invariant under a standard reparametrization
invariance:
\begin {equation}
\delta X ^ \mu ( \sigma _ 1, \sigma  _ 2  , \sigma _ 3
 ) = \epsilon ^ i \partial _ i X ^ \mu ( \sigma _ 1
,\sigma _ 2, \sigma _ 3 )
\end {equation}

The Majorana spinor $\theta$ also transforms as a scalar under
reparametrizations in the world volume variables.4

Under local supersymmetry, we have:
\begin {equation}
\delta X ^ \mu = \bar \theta \Gamma ^ \mu ( 1 + \Gamma )
\kappa ; \,\,\delta \theta = ( 1 + \Gamma ) \kappa
\end {equation}
where $\kappa$ is a local parameter, and:

\begin {equation}
\Gamma = { 1 \over 6 \sqrt { -g } }
\epsilon ^ { ijk}
\Pi _ i ^ \mu
\Pi _ j ^ \nu 
\Pi _ k ^ \rho
\Gamma _ { \mu \nu \rho}
\end {equation}
and where $\Pi _ i \cdot \Pi _ j = g _ { ij }$.

The action as it stands is intractable because of its highly
coupled nature. 
The simplest way of simplifying and quantizing the theory is to
go to the light cone gauge, where all longitudinal modes are
removed. 
We impose:

\begin {equation}
\Gamma ^ + \theta = 0
\end {equation}
along with the usual bosonic constraints.
A large number of terms vanishes in the light cone
gauge because $\bar\theta \Gamma ^ \mu \partial _ i \theta = 0$
except for $\mu = -$.
In particular, the higher order coupled terms of the action
disappear in this gauge.

Then the reduced equations of motion can be derived from the
Hamiltonian:

\begin {equation}
H = \int d ^ 2 \sigma \left [
{ 1\over 2 } ( P ^ I ) ^ 2 +
{ 1\over 4 }
( \{ X ^ I, X ^ J \} ) ^ 2 
- {i \over 2 } \bar \theta \Gamma ^ I \{ X ^ I  , \theta \} \right ]
\end {equation}
where $I = 1,2,...,9$ and:

\begin {equation}
\{ A, B \} = \partial _ 1 A \, \partial _ 2 B - ( 1 \leftrightarrow 2)
\end {equation}
and where the physical states are constrained by:
\begin {equation}
\{ \dot X ^ I , X ^ I \} +  \{ \bar \theta , \theta \} = 0
\end {equation}
which vanishes on physical states.
This constraint generates area preserving diffeomorphisms.

The problem with this Hamiltonian is that, for certain configurations
of the membrane, the potential function, which is the
second term in the Hamiltonian (9), vanishes. This is potentially 
disastrous for the theory. Let $f ( \sigma _ 1, \sigma _ 2 )$
represent a function of the membrane variables, and consider
$ X _ \mu (f )$, which represents a string-like configuration.
For this string-like configuration, the potential function
disappears because:

\begin {equation}
\{ X _ \mu ( f ) , X _ \nu ( f ) \} = 0
\end {equation}

This means that classically, the potential function of the
bosonic Hamiltonian vanishes along string-like filaments with zero area
that protrude from the membrane like the quills of a porcupine.
In principle, this may destabilize the Hamiltonian, allowing
leakage of the wave function along these strings.
In ref. 9, the potential was shown to vanish when 
$X$ was approximated by fields defined in the Cartan sub-algebra
of SU(n). Because the elements of the Cartan sub-algebra
commute among each other, the potential term was shown to
vanish. 

However, it is not obvious that this means that the theory
is unstable along these string-like configurations. Let us
study  a toy-model to understand the subtleties of the question.

As in ref. 12, let us begin with a simple quantum mechanical
system in two dimensions, with the potential given by
$x ^ 2 y ^ 2 $:

\begin {equation}
H _ B = - \Delta + x ^ 2 y ^ 2
\end {equation}

This Hamiltonian resembles the supermembrane theory because the
interaction Hamiltonian vanishes along the $x$ and $y$ axes,
so naively one may expect that the wave function can \lq\lq leak"
along the axes and the theory is therefore unstable.
However, this is not true. Let us temporarily fix the value of
$x$, which is defined to be large. If we move a short distance
along the $y$ axis, the potential function is a potential
well for the harmonic oscillator which is
quite steep for large values of $x$, so the leakage is quite
small. For large $x$, the leakage is infinitesimally small. 
So which effect dominates?

In fact, the spectrum is actually discrete. For fixed $x$,
the Hamiltonian obeys $H  _ B \geq |x|$, so the energy necessary to
move the wave function to infinity is infinite. In fact,

\begin {equation}
H _ B \geq ( |x| + |y|) /2 
\end {equation}
so the spectrum is discrete.

This toy model shows that there are subtleties with regard to the
stability of even simple quantum mechanical systems. However,
the theory can still become unstable if we introduce fermions and
supersymmetry. The zero point energy from the fermions can cancel
the $|x|$ contribution, giving us an unstable theory.

Start with the quantum mechanical system$^9$:
\begin {equation}
H = { 1 \over 2 } \{ Q, Q ^ \dagger \}
\end {equation}
where:

\begin {equation}
Q = Q ^ \dagger = \left (
\begin{array} {cc}
-xy & i \partial _ x + \partial _ y \\
i \partial _ x - \partial _ y &
xy 
\end {array}\right )
\end {equation}

The Hamiltonian reads:

\begin {equation}
H= \left(
\begin {array} {cc}
- \Delta + x ^ 2 y ^ 2 & x + i y    \\
x - i y & - \Delta + x ^ 2 y ^ 2 
\end {array}
\right)
\end {equation}
For fixed $|x|$, the supersymmetry of the reduced system is
enough to guarantee that the energy contribution coming from the
fermionic variables cancels the contribution from the
bosonic variables. In fact, if we define:

\begin {equation}
\xi = { 1 \over \sqrt 2 } 
\left ( 
\begin {array} {r}
1 \\ -1 
\end {array} \right )
\end {equation}
then:
\begin {equation}
\xi ^ T H \xi = H _ B -x
\end {equation}
so the fermionic contribution cancels the $x$ coming from the
bosonic variables, and the system becomes unstable.

We can introduce normalized wave functions for this case as:

\begin {equation}
\Psi _ t ( x , y ) = \chi ( x - t )
\pi ^ { -1/4} |x| ^ { 1/4}
\exp \left ( - { 1 \over 2 }
|x| y ^ 2 \right ) \xi
\end {equation}
for the ground state. $t$ is a parameter which will be taken
to be arbitrarily large; it measures the distance we have
shifted the wave function along the $x$ direction. $\chi$ is
a function which has compact support.
Then we can see that:

\begin {equation}
\lim _ { t \rightarrow \infty} =
( \Psi _ t , H ^ n \Psi _ t ) =
\int dx \, \chi ( x ) ^ * ( - \partial _ x ^ 2 ) ^ n
\chi ( x ) 
\end {equation}
for $n=0,1,2$, so we can shift the wave function as $t$ goes
to infinite without having to supply an infinite amount of
energy. In fact, if $E$ is the energy of this system, we can
see that $E$ can have any arbitrary value,
corresponding to the eigenvalue of $- \partial _ x ^ 2$,
where the potential vanishes. Hence, the energy spectrum
is continuous.

A similar situation may happen with quantum supermembranes. Naively,
the bosonic membrane theory seems to be unstable because the
potential vanishes along certain string-like directions. However, the
amount of energy necessary for the wave function to leak along these
directions is infinite. But when we add fermions into the theory,
then we must check explicitly if the fermionic contribution
to the zero point energy cancels the bosonic contribution.

In ref. 9, this was studied by approximating the membrane with
super Yang-Mills theory. We wish, however, to keep the continuum
limit throughout, and at the very last step identify where any
infinities may arise and where regularization methods may be
necessary.

\section{Zero Point Energy}

Now let us calculate the zero point energy for the quantum supermembrane
in the light cone gauge. Let us divide the original $X ^ I$
membrane co-ordinate into several parts. Let $x$ represent
the co-ordinate along the string, so that:

\begin {equation}
x = x \left ( f ( \sigma _ 1 , \sigma _ 2 ) \right)
\end {equation}

We will let $Y$ be the co-ordinate of the membrane which lies
off the string, i.e. it cannot be written as a function of a
single string variable.

In order to carry out gauge fixing, let us select out the
9th co-ordinate from $I$. Let the $a$ index represent 1,2,...,8.

Now let us split the original $X ^ I $ into different pieces.
Not only will we split the 9th component off from the
others, we will also explicitly split $X ^ I $ into $x ( f ) $
and $Y$.

Then:

\begin {equation}
X ^ I = \left ( x _ 9, Y _ 9 , x _ a , Y _ a \right )
\end {equation}
(At the end of the calculation, we will shift along the
string-like configuration as some parameter
$t\rightarrow \infty$, 
where $ x _ 9 = x (f)$ grows like $t$, while $x _ a$ goes to
a constant. So $x _ a$ can be dropped in relation to $x$, but
we will keep both variables in our equations until the very last
step.)

We can fix the gauge by choosing $Y_9=0$.

Then the Hamiltonian can be split up into several pieces:

\begin {equation}
H = H _ 1 + H _ 2 + H _ 3 +  H _ 4
\end {equation}
where:

\begin {eqnarray}
H _ 1  &= &- { 1 \over 2 }
\int d ^ 2 \sigma \, \left [
\left ( { \partial \over \partial x }\right) ^ 2
+
\left ( {\partial \over \partial x _ a } \right ) ^ 2 \right ]
\nonumber \\
H _ 2 &= &- { 1 \over 2 } \int d ^ 2\sigma \,
\left ( { \partial \over \partial Y _ a }\right) ^ 2 
+
{ 1 \over 2 }
\int d ^ 2 \sigma d ^ 2 \bar \sigma 
d ^ 2 \sigma ' \,
\left [
Y _ a ( \bar \sigma ) z ^ T ( \bar \sigma, \sigma ')
z ( \sigma ' , \sigma ) Y _ a ( \sigma ) \right ]
\nonumber \\
H _ 3 &= &- { i \over 2}
\int d ^ 2 \sigma  d ^ 2 \bar \sigma \,
\bar \theta ( \bar \sigma ) \left [
z ( \bar \sigma, \sigma ) \gamma _ 9 + 
z _ a ( \bar \sigma ,\sigma )\gamma _ a  \right] \theta ( \sigma )
\end {eqnarray}
where:
\begin{eqnarray}
z ( \bar \sigma , \sigma ) &=
&
\delta ^ 2 ( \bar \sigma , \sigma ) \partial _ { \sigma _ 1 }x
\partial _ { \bar\sigma _ 2  } - ( 1 \leftrightarrow 2 )
\nonumber \\
z _ a ( \bar \sigma , \sigma ) &=
&
\delta ^ 2 ( \bar \sigma , \sigma ) \partial _ { \sigma _ 1 } x _ a 
\partial _ { \bar\sigma _ 2  } - ( 1 \leftrightarrow 2 )
\end {eqnarray}
and the index $\sigma$ is shorthand for $\{ \sigma _ 1 , \sigma _ 2 \}$.
Notice that $z(\bar \sigma, \sigma)$ is an anti-symmetric
function. Also, we have set $\Gamma _ a = \gamma ) a$.
$H _ 4 $ contains other terms in $Y$, which will not concern us
yet.

The key factor, which will dominate our entire
discussion of the
zero point energy, is $z (\sigma , \bar sigma )$, which is
the continuous matrix element which defines the diffeomorphism algebra in
equation (10). In particular, we are interested in the sub-algebra
of $w(\infty)$ which defines the reparametrization along the
string-like filament. For elements $x (f )$, the elements of
the algebra commute among each other. (In ref. 9, the counterpart
of $x (f)$ are elements of the Cartan sub-algebra, which
commute among each other by definition.)
$z$ is important to our discussion because the zero
point energy can be defined entirely in terms of its eigenvalues.

Now consider the term $H _2 $. We can write down an eigenfunction
for $H_2$ as:

\begin {equation}
\Phi _ 0 = A ( {\rm det } \, \Omega ) ^ 2 \exp
\left ( - { 1 \over 2 }
\int d ^ 2 \bar \sigma d ^ 2 \sigma \, Y _ a ( \bar \sigma )
\Omega ( \bar \sigma ,\sigma ) Y _ a ( \sigma ) \right)
\end {equation}
where $\Omega$ is yet undetermined, and $A$ is a 
normalization constant, determined by:

\begin {equation}
1 = \left ( \Phi _ 0 ,\Phi _ 0 \right )
=
\int \prod _  a \prod _ \sigma {\cal D } Y _ a ( \sigma ) 
\,\Phi _ 0 ^ * \Phi _ 0 
\end {equation}

Applying $H_2$ to this wave function, we find:

\begin {equation}
H _ 2 \Phi  _ 0 = 4 \int d ^ 2 \sigma \, \Omega ( \sigma , \sigma )
\, \Phi _ 0 
\end {equation}
which fixes the value of $\Omega$ to be:

\begin {equation}
\Omega ^ 2 ( \bar \sigma, \sigma ) = \int d ^ 2 \sigma '
\, z ^ T ( \bar \sigma , \sigma ' ) z ( \sigma ' , \sigma )
\end {equation}

To find an explicit expression for the ground state energy
requires that we take the trace of $\Omega$. This is a tricky
problem, since the trace may actually diverge, requiring a
regularization. Let us assume that we can diagonalize
the $z$ by finding its eigenvalues. Let us
introduce eigenvectors $E _  { MN}^ \sigma$,
where $M \neq N$, as follows:

\begin {equation}
\int d ^ 2 \sigma \, z ( \bar \sigma , \sigma ) 
\,E _ { MN } ^ \sigma = i \lambda _ { MN } E_ { MN } ^ \sigma
\end {equation}
\begin {equation}
\int d ^ 2 \sigma \, z _ a ( \bar \sigma , \sigma ) 
\,E _ { MN } ^ \sigma = i \lambda _ { MN } ^ a E_ { MN } ^ \sigma
\end {equation}
where $M,N$ label a complete set of orthonormal functions,
which can be either continuous or discrete, and $\lambda _ { MN}$
are the anti-symmetric
eigenvalues of $z$. Our discussion will not
depend on the explicit representation. (Since $z$ and
$z _ a $ commute, we can diagonalize them with the
same eigenvectors.)

We can normalize them as follows:

\begin {equation}
\int d ^ 2 \sigma \, ( E _ { MN } ^ \sigma ) ^ *
E _ { PQ } ^ \sigma = \delta _ { MP } \delta _ { NQ }
\end {equation}
\begin {equation}
\sum _ { M\neq N} ( E _ { MN } ^ \sigma ) ^ *
E _ { MN }^ { \bar \sigma } = \delta ^ 2 ( \bar \sigma - \sigma )
\end {equation}
\begin {equation}
( E _ { MN } ^ \sigma ) ^ * = E _ { NM } ^ \sigma
\end {equation}

If we diagonalize $z$ in terms of these eigenvalues, we find that
the eigenvalue of $H _ 2$ is given by the sum of the absolute
values of the eigenvalues of $z$:

\begin {equation}
\int d ^ 2 \sigma \, \Omega ( \sigma , \sigma ) =
\sum _ { M,N } | \lambda _ { MN } |
\end {equation}
\begin {equation}
{\rm det } \, \Omega = \prod _ { M<N} \lambda  _ { MN}^2
\end {equation}
(Because $\lambda _ { MN}$ is anti-symmetric, we can reduce
the product over all $M,N$ to one with $M < N$, where the
precise ordering of the indices is arbitrary.)
Now let us calculate the contribution of the fermionic
variables to the zero point energy.

\section{Fermionic Variables}
The calculation of the fermionic variables is more difficult. As before
our plan is to express all quantities in terms of the eigenvalues
of the matrix $z(\bar \sigma, \sigma )$. Our calculation
will resemble the path taken in ref. 9.

We now change variables to:

\begin {equation}
\theta (\sigma ) = \sum _ { M\neq N} \theta ^ { MN } 
E _ { MN } ^ \sigma
\end {equation}

The original fermionic variables are real. This means, therefore,
that:

\begin {equation}
\theta ^ { MN\dagger} = \theta ^ { NM}
\end {equation}

We can check that the anti-commutation relations:

\begin {equation}
\left [ \theta _ \alpha ( \sigma ) , \theta _  \beta 
( \bar \sigma ) \right ] _ + 
= \delta _ { \alpha ,\beta } \delta ^ 2 ( \sigma - \bar \sigma )
\end {equation}
are transformed into:

\begin {equation}
\left [ \theta _ \alpha ^ { MN } , \theta _ \beta ^ { PQ } 
\right ] _ + = 
\delta _ { \alpha , \beta } \delta ^ { MQ} \delta ^ { NP}
\end {equation}

The fact that we can convert the complex $\theta ^ {MN}$ into
its conjugate by simply reversing the lower indices is quite
convenient, but it will allow us to establish creation and
annihilation operators.

Then $H_3$ reduces to:

\begin {eqnarray}
H _ 3 & = & 
{ 1 \over 2 } \sum _ { M \neq N } \theta  ^ { NM } 
\left ( \lambda _ { MN } \gamma  _ 9 +
\lambda _ { MN } ^ a \gamma _ a \right )  
\theta ^ { MN } 
\nonumber \\
& = & 
\sum _ { M<N} \theta ^ { MN \dagger} 
\left ( \lambda _ { MN } \gamma  _ 9 +
\lambda _ { MN } ^ a \gamma _ a \right )
\theta ^ { MN} 
\end {eqnarray}

Now let us make a change in fermionic variables to eliminate the
presence of $\gamma _ a$ in the above expression. Let us define:

\begin {equation}
\tilde \theta ^ {MN } = ( A _ { MN } +
B _ { MN } ^ a
\gamma _ a \gamma _ 9 ) \theta ^ { MN}
\end {equation}

When we insert this expression back into the one for $H _3$,
we demand that $H _ 3$ reduce down to a function of 
$\tilde \theta ^{MN\dagger} \gamma _ 9
\tilde \theta ^ { MN}$.
We then find a system of two equations:

\begin {eqnarray}
\omega _ { MN }
\left [ A _ { MN } ^ 2 - ( B _ { MN }  ^ { a}  ) ^  2  
\right ] & = &
\lambda _ { MN } 
\nonumber \\
- 2 A _ { MN } B _ { MN } ^ a \omega _ { MN } & = & \lambda _ { MN } ^ a
\end {eqnarray}
whose solutions are
given by:

\begin {eqnarray}
A  _ { MN} & = &
{ 1 \over \sqrt { 2 \omega _ { MN } }}
\sqrt{ \omega _ { MN } +\lambda _ { MN } }
\nonumber \\
B_{ MN } ^ a & = &
- { 1 \over \sqrt { 2 \omega _ { MN } } }
{ \lambda _ { MN } ^ a \over \sqrt { \omega _ { MN }+\lambda _ { MN } } }
\nonumber \\
\omega _ { MN } & =  & \sqrt { \lambda _ { MN } ^ 2 
+ ( \lambda _ { MN }^ { a} ) ^  2  }
\end {eqnarray}

So the expression for $H _ 3$ reduces to:

\begin {equation}
H _ 3 = \sum _ { M,N} \omega _ { MN }
\tilde \theta ^ { MN \dagger}
\gamma _ 9 \tilde \theta ^ { MN }
\end {equation}

Lastly, in order to eliminate the presence of $\gamma _ 9$,
let us introduce projection operators:

\begin {equation}
P _ \pm = { 1 \pm \gamma _ 9 \over 2 } 
\end {equation}
so that:
\begin {equation}
\phi \gamma _ 9 \theta = \phi _ + \theta _ + 
- \phi _ - \theta _ -
\end {equation}

Then $H _ 3$ becomes:

\begin {eqnarray}
H _ 3 & = & 
\sum _ { M < N }
\left ( 
\tilde \theta _ + ^ { MN\dagger }
\tilde \theta _ + ^ { MN } 
- \tilde
\theta _ - ^ { MN \dagger }
\tilde \theta _ - ^ { MN } \right )
\nonumber \\
& = &
\sum _ { M<N }
\left ( 
\tilde \theta _ + ^ { MN\dagger }
\tilde \theta _ + ^ { MN } 
+\tilde
\theta _ - ^ { MN }
\tilde \theta _ - ^ { MN \dagger } - 8 \right )
\end {eqnarray}

We are interested in the last constant in order to calculate
the ground state energy of the system.

\section {Wave Function}

We can now write down the wave function. 
Since $\theta  ^ { MN}$ is the conjugate to
$\theta ^ { NM}$ for $M<N$, we can choose $\theta ^ { MN}$
to be annihilation operators.
Let $\xi _ 0$ represent
the vacuum state vector such that the annihilation operators act as
follows:

\begin {equation}
\theta ^ { MN } \xi _ 0 = 0
\end{equation}
for all indices such that $M< N$. Then the ground state
vector for the fermionic variables is:

\begin {equation}
\xi = \left[
\prod  _  {M<N} \prod _  \alpha ^ 8 
\left ( 
\tilde \theta _ - ^ { MN \dagger } 
\right ) \right ] 
\xi _ 0 
\end {equation}
In particular, this means that:

\begin {eqnarray}
\tilde \theta _ + ^ { MN } \xi &= &0
\nonumber \\
\tilde \theta _ - ^ { MN\dagger} \xi &= &0
\end {eqnarray}
With this choice, we see that:

\begin {equation}
H _ 3 \xi = - 8 \sum _ { M<N} \omega _ { MN }\xi
\end {equation}

The total wave function can now written as:

\begin{equation}
\Psi = \chi (x - t V, x _ a ) 
\Phi _ 0 ( x , Y _ a ) \xi ( x , x _ a )
\end{equation}
where $t$ becomes large as we go along the string, and $V$ represents
the asymptotic value of the string variables, which depends on the
function $f$.

To find total energy, we now sum the contribution of
$H _ 2$ and $H _3$:

\begin {equation}
( H _ 2 + H _ 3 ) \Psi = 8
\sum _ { M< N} \left ( 
|\lambda _ { MN }| - \omega _ { MN } \right ) \Psi
\end {equation}

As before, let $t$ represent a variable which measures how far
we are along the string-like filament. We shall take $t\rightarrow
\infty$. We make the assumption that $x$ grows as $t$, while $x _ a$
approaches a constant. Then we see that $\omega _ { MN}$
asymptotically approaches $|\lambda _ { MN }|$ in this limit,
so that the ground state energy of $H _ 2$ and $H _ 3$
vanishes:

\begin {equation}
\left ( H _ 2 + H _ 3 \right) \Psi \rightarrow 0
\end {equation}

It is not hard to find the contribution of $H _ 4$, which is a
polynomial in $Y,x,x _ a$.  We are interested in the value
of the matrix element:

\begin {equation}
\lim _ { t \rightarrow \infty} | \Psi , P \Psi | \rightarrow
t ^ n
\end {equation}
for some polynomial $P$. Since $\Phi_ 0 $ is just the
ground state wave function for the harmonic
oscillator in terms of $Y$, 
it is easy to calculate the value of 
$(\Phi _ 0 , P(Y) \Phi _ 0)$. We find that 
$n = -1/2$ for every $Y$ contained within $P$. 
For every $x$ contained within $P$, we have a contribution
of $n=1$. By simply counting the number of $x$
and $Y$ within $H _ 4$, we see that it does not contribute to the
ground state energy to the leading order, so it can be ignored.

In conclusion, we see that the principal contribution to the
ground state energy comes from $ H _ 2 $ and $H _ 3 $.

Furthermore, we see that the energy eigenvalue of the operator
is continuous for the ground state, which means that the system is
unstable.

We caution that there may be hidden infinities with regard to
our calculation. Since the continuous matrix $z( \sigma ,\bar \sigma)$
contains derivatives, it may be possible that its eigenvalues are
actually divergent. Then the cancellation of the lowest eigenvalue of
$H _ 2 + H _ 3$ must be carefully regularized. However, the
advantage of our discussion is that it was carried out in the
continuum theory rather in super Yang-Mills theory, so 
we have a more intuitive understanding of
where the problems may arise. 

\section {Discussion}

Although the system is unstable, we speculate how this may still
be compatible with known phenomenology. 
For a physical system like quantum membranes to be compatible with
known physics, we have to ask:

\noindent a) why don't we see them in nature?

\noindent b) do decaying fundamental particles violate unitarity
or other cherished features of quantum field theory?

To answer the first objection, we note that because 
the decay time of such a quantum membrane is on the order of
the Planck time, it is possible that unstable membranes decay
too rapidly to be detected by our instruments. 

One potential flaw in this argument is that there may be
different values of physical parameters, such as the mass of
the membrane, for which the life-time is long.
Therefore, to 
make a rough guess of the decay life of such a quantum membrane,
we recall that the decay width of the decay of a quark-anti-quark
bound state is given by:

\begin {equation}
\Gamma = { 16 \pi \alpha ^ 2 \over 3 }
{ \left | \psi ( 0 ) \right | ^ 2 \over
M ^ 2 }
\end {equation}
where $\phi (0 )$ is the wave function of the bound state at the origin,
and $M$ is the mass of the decay product. On dimensional 
and kinematic grounds, we expect this formula to be roughly correct
for the decay of the membrane, discarding the effect of spin, quantum
numbers, etc. 

We expect that $| \psi ( 0 )|$ to be on the order of a ${\rm fermi}^ {-3}$,
the rough size of the quark-anti-quark bound state. For our purposes,
we assume that the membrane is on the order of the Planck length.
On dimensional grounds, we therefore expect that 
the lifetime of the membrane to be on the order of:
\begin {equation}
T \sim { M ^ 2 L ^ { -3} }
\end {equation}
where $L$ is the Planck length.

For relatively light-weight membranes, we find that the 
lifetime is much smaller than Planck times, so we will, as expected,
never see these particles. 

The other case is more interesting. For very massive membranes,
we find that the lifetime becomes arbitrarily long, which seems
to violate experiment. However, the coupling of very massive membranes,
much heavier than the Planck mass,
is very small, and hence barely couple 
to the particles we see in nature.
Again, we see that unstable membranes cannot be measured in the laboratory.

In summary, light-weight membranes live too short to be detected,
and massive (long-lived) membranes have vanishing coupling
to the known particles.

Yet another objection that one may raise to our naive
arguments is that
membranes decay into other membranes, losing energy and mass with
each decay, and hence the lifetime of the decaying membranes
constantly changing. It is conceivable that, starting with a
single membrane, the cascading sequence of daughter membranes
may produce a collection of membranes 
with a lifetime long enough to be 
measured in the lab. 

To estimate the effect of an infinite sequence of decaying membranes,
let us analyze a simpler system: a chain of 
decaying objects, similar to the decay of a series of radio-nuclides.

Let $N_i$ represent the amount of decaying material of type $i$.
Let $\Omega _ i$ represent the rate of decay of the $i$th  material.
Let $\omega _ { ij}$ represent the rate at which substance
$i$ is decaying into substance $j$, which increases the 
amount of the $j$th substance.
Then the coupled series of equations is given by:

\begin {eqnarray}
\dot N _ 1 &= & - \Omega _ 1 N _ 1 
\nonumber \\
\dot N _ 2 & = & - \Omega _ 2 N _ 2 
+ \omega _ { 12 } N _ 1
\nonumber \\
\dot N _ 3 & = & 
-\Omega _ 3 N _ 3 + 
\omega _ { 13} N _ 1
+ \omega _ { 23 } N _ 2
\nonumber \\
&\vdots &\nonumber\\
\dot N _ j & = & 
-\Omega _ j N _ j =
\sum _ { k=1} ^ { j-1} \omega _ { kj } N _ k
\end {eqnarray}

There is a simple solution to these coupled equations.
If we use the ansatz:

\begin {eqnarray}
N _ 1 & = & A _ 1 e ^ { -\Omega _ 1 t }
\nonumber \\
N _ 2 & = & A _ 2 e ^ { -  \Omega _ 2 t } +
A_ { 12 } e ^ { -\Omega _ 1 t }
\nonumber \\
N _ 3 & = & A _ 3 e ^ { -\Omega _ 3 t }
+ A _ { 23} e ^ { -\Omega _ 2 t } 
+ A _ { 13 } e ^ {-\Omega _ 1 t }
\nonumber \\
& \vdots &
\nonumber \\
N _ j & = & A _ j e ^ { - \Omega _ j t }
+ \sum _ { i = 1} ^ {j-1} A _ { ij } e ^ { -\Omega _ i t }
\end {eqnarray}
then the solution is given by:

\begin {eqnarray}
A_ { 12 } & = & { \omega _ { 12 } A _ 1 \over \Omega _ 2 - 
\Omega  1 }
\nonumber \\
A _ { 13 } & = & { \omega _ { 13 } A _ 1 \over
\Omega _ 3 - \Omega _ 1 }
+
{\omega _ { 23 } 
\omega _ { 12 } A _ 1 
\over
(\Omega _ 2 - \Omega _ 1 ) ( \Omega _ 3 -\Omega _ 1 ) }
\nonumber \\
A _ { 23 } & = & 
{\omega _ { 23 } \omega _ { 12 }
\over  (\Omega _ 3 - \Omega _ 2 )( \Omega _ 2 - \Omega _ 1 ) } A _ 2 
\end {eqnarray}
and so on.

The lesson we learn from this is that, even with an arbitrarily large
number of decaying products, each decaying into each other, the
limiting factor is the longest life-time of a single decay product.
The substance with the slowest decay $ e ^ { - \Omega _ i t }$
dominates the entire series.

This gives us reasonable assurance that the infinite series of decaying
membranes does not behave any worse than its longest lived membrane.

And lastly, we must ask the question of whether quantum field theory
can accomodate decaying fundamental particles.
Previous work by McCoy and Wu$^{13,14}$ on the field theory of
decaying particles indicates that there are no fatal problems
with such a theory. The pole of the two-point function
corresponding to an unstable
particle becomes a branch cut in this case. The principle
question is whether there exists a Lehmann spectral representation
of such a theory with a branch cut.
In fact, two-point functions with a branch cut rather than
a single pole have already been encountered in two
dimensional SU(N) Yang-Mills theory in the limit $N\rightarrow \infty$,
the two-dimensional SU(2) Thirring model, and two-dimensional
Ising field theory.

Of course, these arguments that we have presented here are 
certainly not rigorous, since we do not know how membranes
interact and the theory is highly non-linear. 
Until the interacting theory is fully calculated, we cannot
know precisely whether our heuristic arguments will hold up.
However, they indicate that one cannot
immediately dismiss fundamental quantum membranes as a physical theory.

\end{document}